# Movable Gate MOSFETs as Readout Devices for Cantilever-Based Nano-Electromechanical Sensors

Z. Geng[2*], A. Hessel[1,*], S. C. Scholz[1], F. Schwierz[2], M. Ziegler[3], and J. Knoch[1]

[1]*Institute of Semiconductor Electronics, RWTH Aachen University*

[2]*Department of Micro- and Nanoelectronic Systems, Technische Universität Ilmenau*

[3]*Department of Materials Science, Christian-Albrechts-Universität Kiel*

**These authors contributed equally to the manuscript*

**Abstract**

This work proposes a novel readout mechanism for highly sensitive cantilever-based mass sensors using movable-gate (MG) MOSFETs. Traditional dynamic-mode cantilever sensors measure mass through shifts in resonance frequency, which offer high precision but require complex analog circuitry and large device areas, limiting integration and miniaturization. In contrast, the proposed approach exploits strong short-channel effects in MG MOSFETs, where the drain current depends exponentially on cantilever position, allowing for orders-of-magnitude changes without analog-to-digital conversion. Numerical simulations and initial experiments demonstrate the feasibility of this approach, which also avoids pull-in instability by using the electrostatic behavior of the gate-channel system to stabilize and even excite oscillations. The concept paves the way for scalable, low-complexity, high-resolution mass sensing with full circuit integration potential.

## 1. Introduction

Nano-electromechanical systems (NEMS) sensors have attracted a great deal of interest and are considered a key technology for applications such as biosensors, artificial noses, etc. [1-5]. In particular, cantilever-based sensors are of great interest, as converting the target event into a change in mechanical properties or state which enables the detection of a wide range of measurands — from the mass of individual molecules [6–7] and the presence of trace elements [8] to environmental factors such as sound, pressure, temperature, and beyond [9,10]. A major benefit of cantilever sensors is their easily scalable geometry, allowing for an adjustment of the sensitivity in a beneficially wide range. Moreover, employing mature silicon fabrication technology enables manufacturing cantilever sensors in a reproducible way and holds promise to co-integrate NEMS sensors alongside digital and analog circuitry for signal processing on the same chip [11-13].

Cantilever sensors can be used in a static and a dynamic mode. In both cases, the position of the cantilever needs to be detected for encoding sensorial information. This can, for instance, be accomplished by integrating a transistor into a cantilever-based NEMS sensor as demonstrated recently [14-16]. Employing the cantilever as a movable gate electrode allows to use the distance between the gate and the channel of a MOSFET (metal-oxide-semiconductor field-effect transistor) to directly encode the actual measurement variable into an electrical signal. For example, the adsorption of molecules, or pressure and temperature differences on the cantilever result in strain that deflects the cantilever and changes the distance between gate and channel. However, in the static mode a substantial change on the functionalized layer of the cantilever, such as a large enough amount of molecules needs to be absorbed, to lead to a measurable deflection. Since this deflection is usually very small, an optical detection mechanism – similar to that of a scanning force microscope – would have to be used, which is bulky, cannot be integrated, and hence restricts the measurement setup to consist of a few cantilever sensors only [3, 8, 17-19]. In addition, detecting a static deflection of the cantilever makes the sensor prone to noise and unwanted impact from the environment, mainly causing a position dependence and requiring a vibration-free measurement environment; both of which render this approach unsuitable for an in-field use.

A dynamic mode is advantageous here, in which an actuator can be used to induce mechanical oscillations of the cantilever. In this case, the shift in mechanical resonance frequency, for example upon adsorption of a molecule (or particle), enables the device to exhibit its favorable properties as a highly sensitive and accurate weight scale [6-7]. Moreover, if a shift of the resonance frequency is used as the readout mechanism, a substantially more stable system is obtained which can – in principle – detect very small changes of the mass of the cantilever, independent of static forces and superimposed noise. Despite this advantage, the read-out of the cantilever's resonance frequency then itself becomes the drawback as it requires an actuator with tunable frequency and a rather complex circuitry to detect the resonance frequency and its shift. Moreover, if the frequency of the cantilever oscillation is detected based on a change of the MOS capacitance, a sufficiently large area is required that in turn prohibits downscaling of the cantilevers and integrating a larger number of cantilevers onto a sensor chip.

To avoid large integration areas and complex readout electronics, a readout mechanism relying on the modulation of the on-state current of a MOSFET has been suggested, see, e.g., [16], However, this approach only yields a $1/t_{\text{G-ox}}$ ($t_{\text{G-ox}}=t_{\text{air}}$ is the gate oxide thickness, see Figure 2) dependence of the drain current $I_D$ and, in addition, bears the possibility of a so-called 'pull-in' of the cantilever when the distance between cantilever (which acts as the transistor's gate) and channel becomes too small [20]. Furthermore, in all cases mentioned so far, analog-to-digital signal conversion is necessary to further process the measured data.

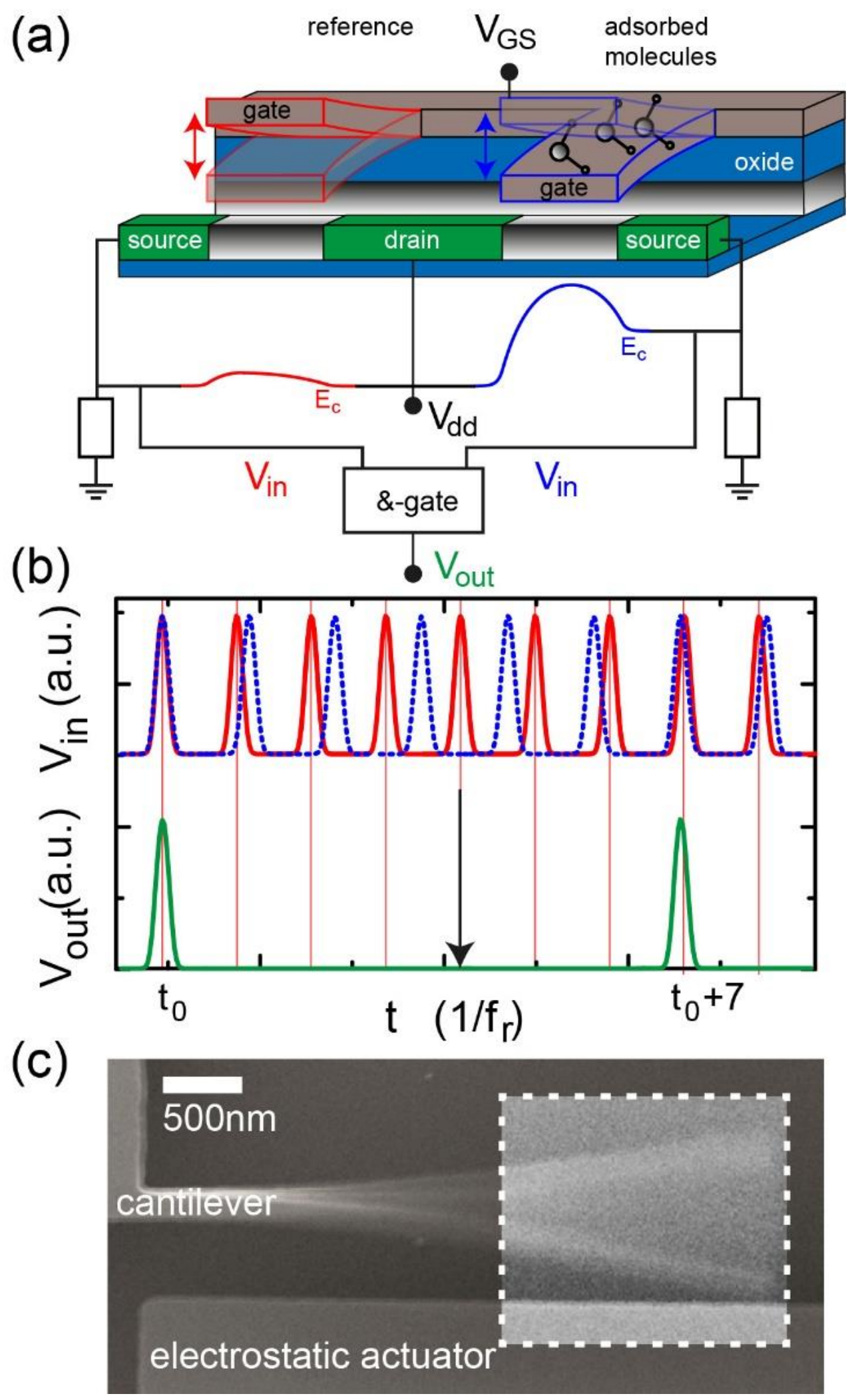

**Fig. 1.** (a) Schematic of two moving gate (MG) cantilever MOSFETs. The modulation of short channel effects (two conduction band profiles $E_c$ are depicted underneath the transistors) is exploited to realize orders of magnitude difference in drive current depending on the position $y$ of the cantilevers. (b) When excited in self-sustained oscillations with resonance frequency $f_{r,1}$ and $f_{r,2}$, the beating between the two MG cantilever MOSFETs can be used to extract e.g. the weight of an adsorbed molecule by counting clock-cycles (employing a clock frequency larger than $f_{r,1}$) in between two subsequent AND events. The black arrow shows the time when the two cantilevers oscillate with 180° phase shift as depicted in (a). (c) Self-sustained oscillations of a Si cantilever induced by applying a constant voltage between cantilever and electrostatic actuator.

In the present paper we introduce a novel sensor consisting of a field-effect transistor (FET) with a movable cantilever as gate electrode that exploits the modulation of short channel effects as a function of cantilever movement. The outstanding features of our sensor are a read-out current that exponentially depends on the position of the cantilever with respect to the FET. Second, in contrast to existing cantilever sensors, in the present design the Coulomb force *drops* when the cantilever approaches the FET thus preventing a pull-in of the cantilever. Both features allow to integrate cantilever MOSFETs into a sensor system with two oscillating cantilevers as depicted in Fig. 1 (a) and use their phase relationship for highly precise mass detection. By connecting the two cantilever FETs to a CMOS AND gate, smallest mass changes can be detected by counting clock cycles at the beat frequency – without the need for analog signal processing. Furthermore, when the cantilevers can be stimulated to oscillate by only applying constant voltages; an experimental example is shown in Fig. 1 (c). Hence, our sensor concept provides a simple, robust highly-sensitive and miniaturizable readout method. Here, we show the results of numerical device simulations of MG cantilever MOSFETs as well as experimental results regarding the proposed readout scheme.

## 2. Theoretical Investigation of the Behavior of MG MOSFETs

### *A. Method*

Figure 2 shows schematically the considered MG MOSFET structure based on silicon-on-insulator (SOI), in which the cantilever acts as movable gate. Strictly speaking, this device is not a MOSFET but rather a MISFET (metal-insulator-semiconductor FET) since here the gate dielectric is not an oxide but the insulating air gap. Nevertheless, we keep the more common term MOSFET. The width of the cantilever is equal to the gate length $L_G$ of the transistor and the air gap of thickness $t_{air}$ represents the gate dielectric.

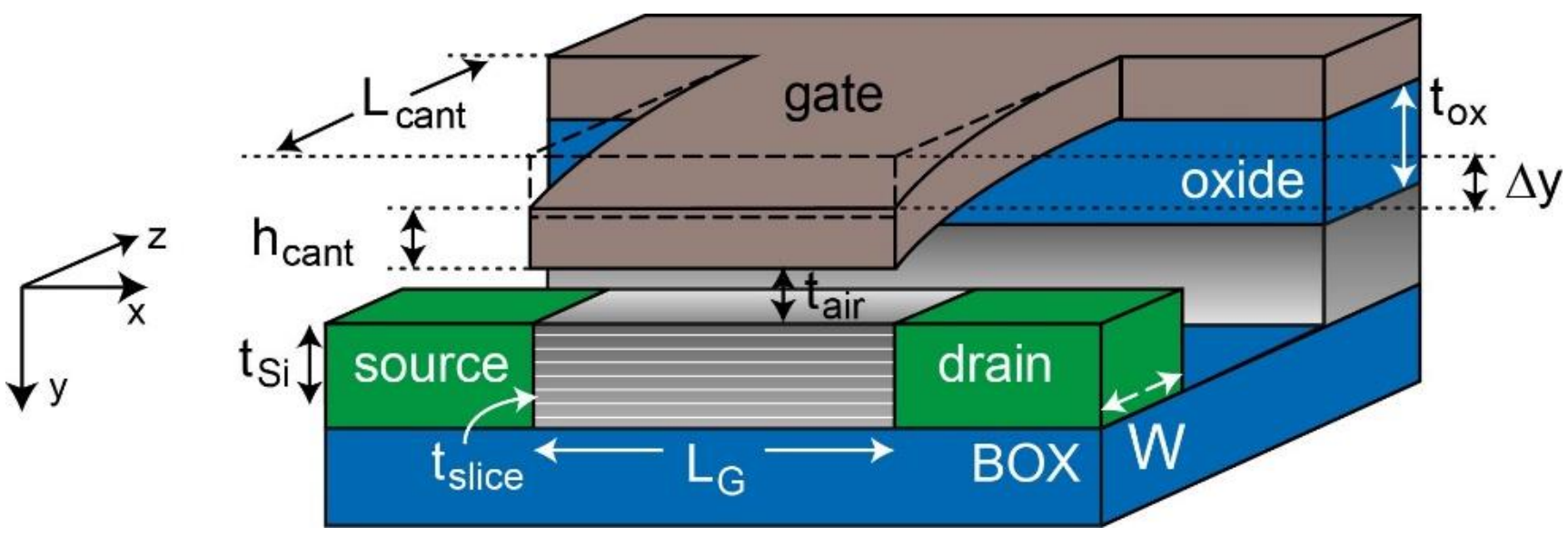

**Fig. 2.** Schematic of a n-channel MG SOI MOSFET consisting of a cantilever (brown) acting as gate and air serving as gate dielectric, heavily doped source and drain regions (green) with the body (grey) having a thickness $t_{Si}$ in between, and the BOX (buried oxide, blue) . The cantilever has the length $L_{cant}$, the height $h_{cant}$, and a width equal to the gate length $L_G$ of the MOSFET and is anchored on an oxide layer (blue) with a thickness $t_{ox}$. Also shown are the deflection of the cantilever tip $\Delta y$ and the coordinate system indicating the spatial directions $x$, $y$, and $z$.

Applying a gate-source voltage $V_{GS}$ to the device causes a certain net charge in the channel and a countercharge of the same magnitude but opposite polarity at the gate. This, in turn, results in a Coulomb force and a corresponding bending of the cantilever gate, which is largest at the cantilever tip and zero at the anchorage point located on top of an oxide layer. Thus, the air gap thickness $t_{air}$ varies along the cantilever length $L_{cant}$, in other words, along the channel width. This represents a complex problem for the theoretical investigation of the operation of MG MOSFETs, the solution of which would require multiple 3D (three-dimensional) device simulations in a loop resulting in a high computational effort. The loops are necessary since the first device simulation (first loop) should be conducted for the case of zero deflection, i.e., $t_{air} = t_{ox}$. From the simulation results, the net charge distribution in the body can be extracted, which in turn can be used to calculate the Coulomb force and the resulting $z$-dependent cantilever deflection. From the latter, an updated $z$-dependent $t_{air}$ is obtained, that has to be used in the simulation during the second run through the loop, which again leads to a new updated $y$-dependent $t_{air}$. Thus, successive 3D simulations with continually updated $t_{air}$ ($z$) would have to be performed until a steady state is reached.

To avoid multiple time-consuming 3D simulations and to keep the computational expenses within reasonable limits, we make a number of simplifying assumptions and apply a stripped-down analysis approach whose key steps are shown in Fig. 3 and will be explained in the following.

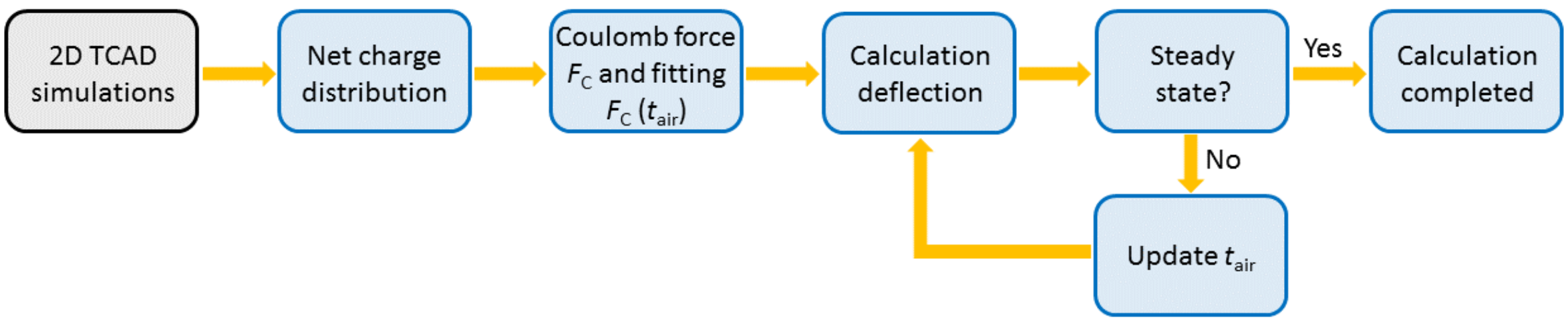

**Fig. 3.** Key steps of the approach for analyzing the behavior of MG MOSFETs. Grey box: Numerical device simulation. Blue boxes: Post-processing.

The analysis starts with 2D drift-diffusion simulations of MG MOSFETs with $N$ different $z$-independent air gap thicknesses and otherwise identical device designs using the commercial TCAD tool Silvaco Victory [21]. Then, as first step of the post-processing procedure, the channel is divided into $m$ stacked thin slices, each with a thickness $t_{slice} = t_{Si}/m$ (illustrated with the white lines in Fig. 2), and the net charge $Q_{net\text{-}i}$ ($i$ is the number of the slice, see Fig. 2) in each of these slices is calculated. Next, the Coulomb force is to be computed. Before this can be done, however, the following consideration has to be made. The net charges in the slices "see" a layered medium with two layers, one consisting of Si and the other of air, between themselves and the gate. For the net charge in a slice close to the bottom of the body, the Si layer is much thicker than for the net charge in a slice close to the body top. This is taken into account by calculating an equivalent dielectric constant $\varepsilon_{eq\text{-}i}$ of the layered medium between the net charge in the $i^{th}$ slice and the gate by considering an equivalent circuit of two capacitors connected in series. The first capacitor accounts for the capacitance of the air gap and the second for the capacitance occurring between the body surface and the $i^{th}$ slice. The equivalent dielectric constant related to the $i_{th}$ slice is then given by

$$\varepsilon_{eq\text{-}i} = \frac{\varepsilon_{air}\, \varepsilon_{Si} \left(t_{air} + t_{Si\text{-}i}\right)}{\varepsilon_{air}\, t_{Si\text{-}i} + \varepsilon_{Si}\, t_{air}} \qquad (1)$$

where $\varepsilon_{air}$ and $\varepsilon_{Si}$ are the dielectric constants of air and silicon and $t_{Si\text{-}i}$ is the distance between the $i^{th}$ slice and the body top. To avoid confusion, note the different meanings of the symbols $t_{Si}$ (body thickness) and $t_{Si\text{-}i}$.

Now, the partial Coulomb force $F_{\text{C-i}}$ originating from the net charge in the $i^{\text{th}}$ slice is calculated and the overall Coulomb force $F_{\text{C}}$ is obtained by adding the Coulomb forces of all individual slices by

$$F_{\text{C-i}} = \frac{Q_{\text{net-i}}^2}{4\,\pi\,\varepsilon_{\text{eq-i}}\left(t_{\text{Si-i}} + t_{\text{air}}\right)^2} \qquad F_{\text{C}} = \sum_i F_{\text{C-i}} \tag{2}$$

The net charges in the channel slices, the equivalent dielectric constant as well as the Coulomb forces $F_{\text{C-i}}$ and $F_{\text{C}}$ are calculated for the different $N$ air gap thicknesses mentioned above and fits for the $F_{\text{C}}$ ($t_{\text{air}}$) dependency are elaborated.

Finally, the resulting deflection $\Delta y$ of the cantilever gate at position $z$ can computed by

$$\Delta y\left(z\right) = \frac{F_{\text{C}}\,L_{\text{cant}}^2\left(6\,L_{\text{cant}}^2 - 4\,L_{\text{cant}}z + z^2\right)}{24\,E\,I} \tag{3}$$

assuming a uniform Coulomb force along the cantilever length [22]. In (3), $E$ is Young's modulus of the cantilever material and $I$ is the moment of inertia of the cantilever given by $L_{\text{G}} \times h_{\text{cant}}^3/12$ [22]. For sake of simplicity, we consider only the deflection at $z = L_{\text{cant}}$, i.e., at the cantilever tip. As shown in Fig. 3, the calculation of the deflection is followed by a check Steady State reached? Depending on the outcome of the check, either the loop has to be run through again since a steady state is not reached yet or the computation is completed (steady state reached). Some more details of the loop are explained in the following.

After calculating the deflection at the cantilever tip for the first time, we change (update) $t_{\text{air}}$ in a way that the new $t_{\text{air_new}} = t_{\text{air_old}} - \Delta y$, i.e. $t_{\text{air}}$ in the first calculation ( $= t_{\text{ox}}$) minus the deflection $\Delta y_1$ at the cantilever tip. We now assume that the entire cantilever is moved closer to the body surface by $\Delta y_1$, and for this condition using the fit $F_{\text{C}}(t_{\text{air}})$ a new Coulomb force $F_{\text{C2}}$ and a new deflection $\Delta y_2$ are calculated. The loop is run through until in two consecutive runs identical new $t_{\text{air}}$ are obtained (i.e., steady state reached) or the cantilever tip touches the body surface which is the so-called the pull-in effect mentioned above. Assuming that, as discussed above, the entire cantilever is moved closer to the body is equivalent to assuming a cantilever with very large $L_{\text{cant}}$ which is of course a rough approximation. It is justified, however, by the aim of our study at the present stage, which is to develop a basic understanding of the operation of MG MOSFETs. In particular, we aim at determining the order of magnitude of the Coulomb force and the resulting deflection of the cantilever.

***B. Results***

The approach for analyzing the behavior of MG MOSFETs described above is now applied to an exemplary model device with the design shown in Fig. 2. We consider a 50-nm gate SOI nMOSFET having an $n^+$ polysilicon cantilever gate with a doping of $10^{20}$ cm$^{-3}$, a length $L_{\text{cant}}$ of 1 µm and a height $h_{\text{cant}}$ of 10 nm and air as gate dielectric. The body thickness $t_{\text{Si}}$ is 10 nm, the body doping is either zero (i.e., undoped body, called structure 1), $10^{17}$ cm$^{-3}$ (referred to as structure 2), or $5\text{x}10^{18}$ cm$^{-3}$ (i.e. structure 3), and the doping of the $n^+$ source and drain regions is $10^{20}$ $cm^{-3}$. The oxide on top of which the polysilicon cantilever is anchored has a thickness of 30 nm and for the Young's modulus of the cantilever a value of 170 GPa is assumed [23].

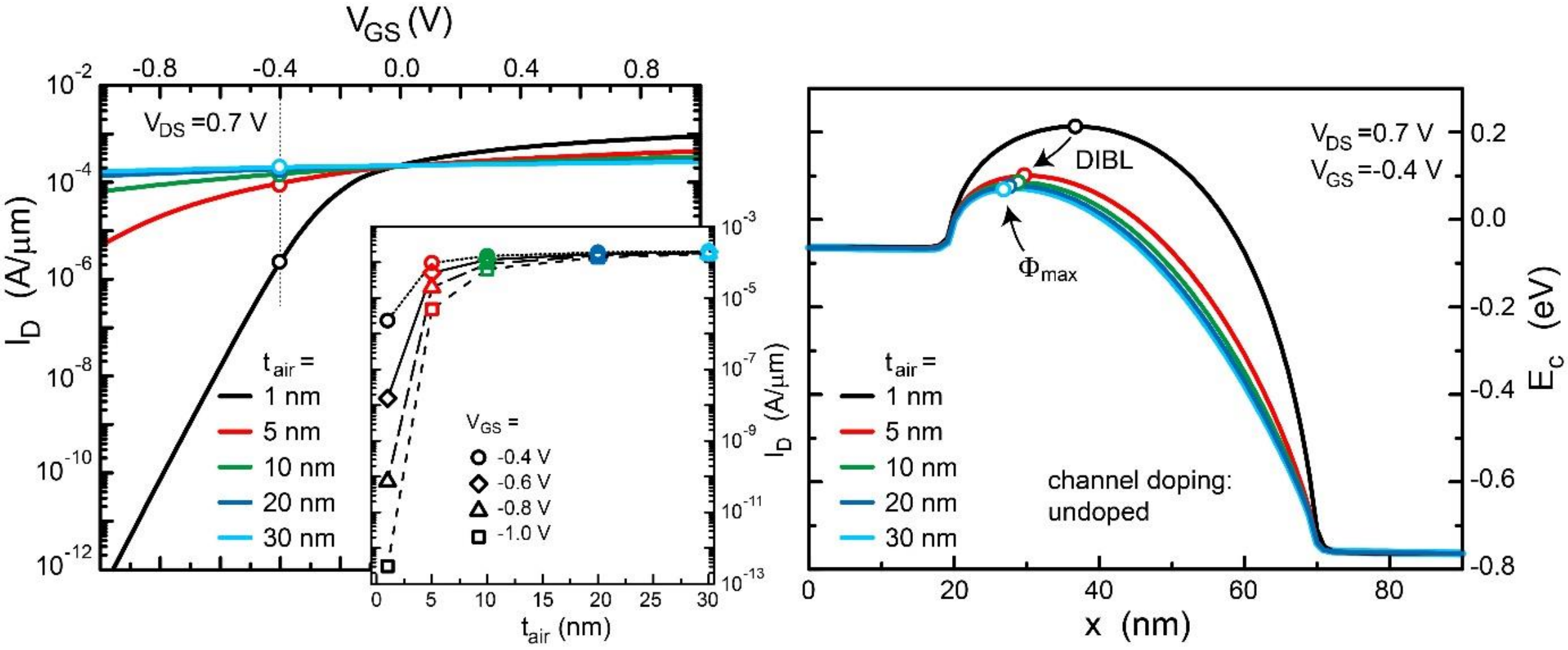

**Fig. 4.** Transfer characteristics (left) and conduction band profiles (right, at $V_{GS}$=-0.4 V) of the model transistor, structure 1 (for a constant drain-source voltage of 0.7 V) for different air gap thicknesses $t_{air}$ in nm. The inset shows the drain current extracted at different constant $V_{GS}$ as a function of $t_{air}$.

The left panel of Fig. 4 shows the transfer characteristics of structure 1 for different air gap thicknesses varying between 1 nm and 30 nm. As to be expected, $t_{air}$ has a strong influence on the shape of the transfer characteristics and hence on the value of the drain current $I_D$ at a given, constant gate-source voltage. For thicker $t_{air}$, short-channel effects become stronger and deteriorate the characteristics to an increasing extent. For instance, in the off-state at $V_{GS}$ = -0.4, the drain current is $7.6 \times 10^{-8}$ A/µm and the subthreshold swing amounts to ~86 mV/dec for $t_{air}$ = 1 nm compared to $4.1 \times 10^{-5}$ A/µm and 3,600 mV/dec for $t_{air}$ = 30 nm. The inset shows $I_D$ as a function of $t_{air}$ extracted for four different $V_{GS}$ in the off-state of the device. A strong dependence with several orders of magnitude change of $I_D$ is observed. This behavior is explained in Fig. 4, right panel. Here, conduction band profiles at $V_{GS}$ = -0.4 V for different $t_{air}$ are shown. Obviously, the conduction band maximum $\Phi_{max}$ (illustrated with circles) that determines the injection of carriers and hence the drain current strongly depends on $t_{air}$. As a result, even for gate voltages within the off-state of the device, large currents are obtained for thicker $t_{air}$ because of the strong leakage due to the drain-induced-barrier lowering (DIBL). We note that for MOSFETs to be used in digital CMOS circuits short-channel effects as strong as those for the model transistor with 10, 20, and 30 nm air gap thickness are in no way acceptable. In our MG MOSFET, on the other hand, strong short-channel effects are exploited and therefore beneficial: they lead to the strong dependence of the drain current as a function of $t_{air}$ shown in the inset of Fig. 4.

A similar behavior as discussed so far is observed for structure 2, as displayed in Fig. 5. However, while the channel doping of $10^{17}$ cm$^{-3}$ and a body thickness of $t_{Si}$=10 nm yields a fully-depleted device with a nearly unchanged inverse subthreshold swing of 86 mV/dec at $t_{air}$ = 1 nm (black curves in Fig. 4 and 5), the doping leads to a shift of the threshold voltage (and thus a larger $\Phi_{max}$ at $t_{air}$ = 1 nm and $V_{GS}$ = -0.4 V) and at larger $t_{air}$ results in a small contribution of a depletion capacitance $C_{depl}$ which is in series with the oxide (=air) capacitance. As a result, a somewhat more gradual increase of drain current for increasing $t_{air}$ is obtained as shown in the inset of Fig. 5, left panel. Hence, doping the channel together with choosing an appropriate $V_{GS}$ enables tuning the transduction of a mechanical movement into an electrical signal. The impact of channel doping will be further discussed below.

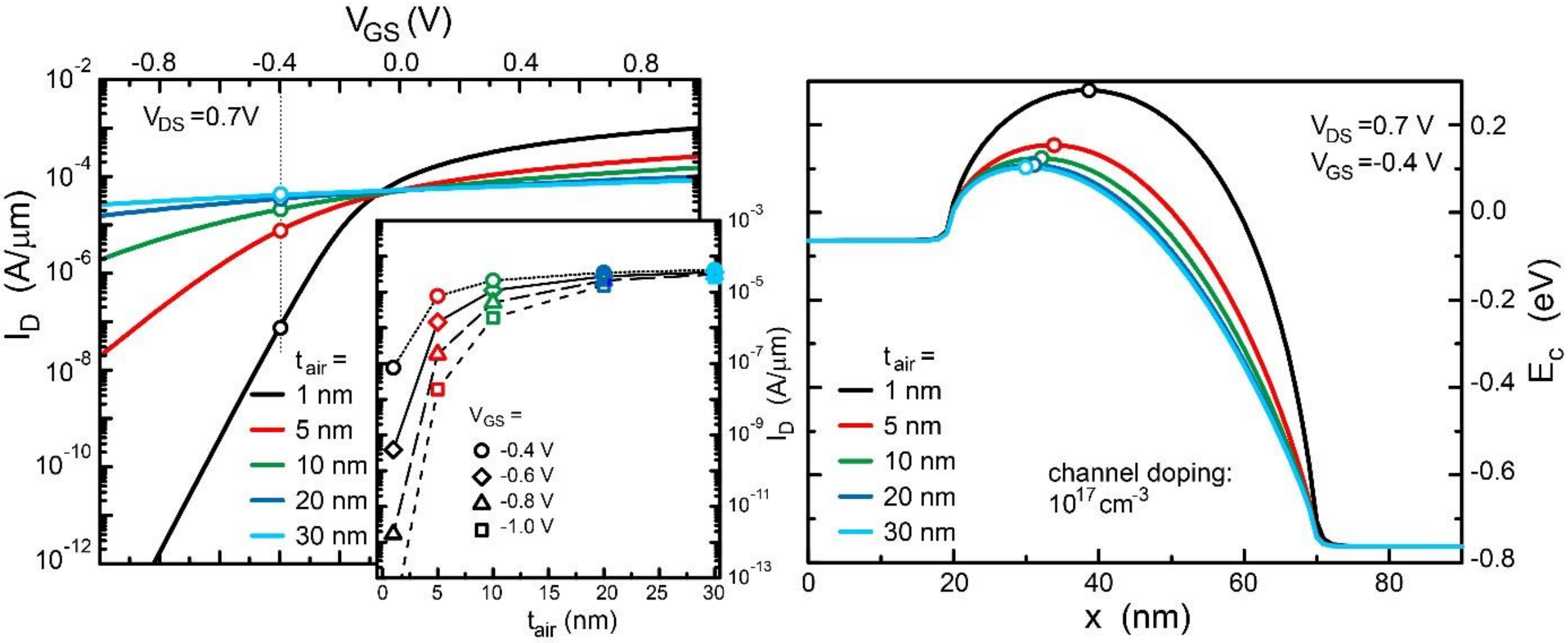

**Fig. 5.** Transfer characteristics (left) and conduction band profiles (right, at $V_{GS}$=-0.4 V) of the model transistor, structure 2 with channel doping $10^{17}$ cm$^{-3}$ (for a constant drain-source voltage of 0.7 V) for different air gap thicknesses $t_{air}$ in nm. The inset shows the drain current extracted at different constant $V_{GS}$ as a function of $t_{air}$.

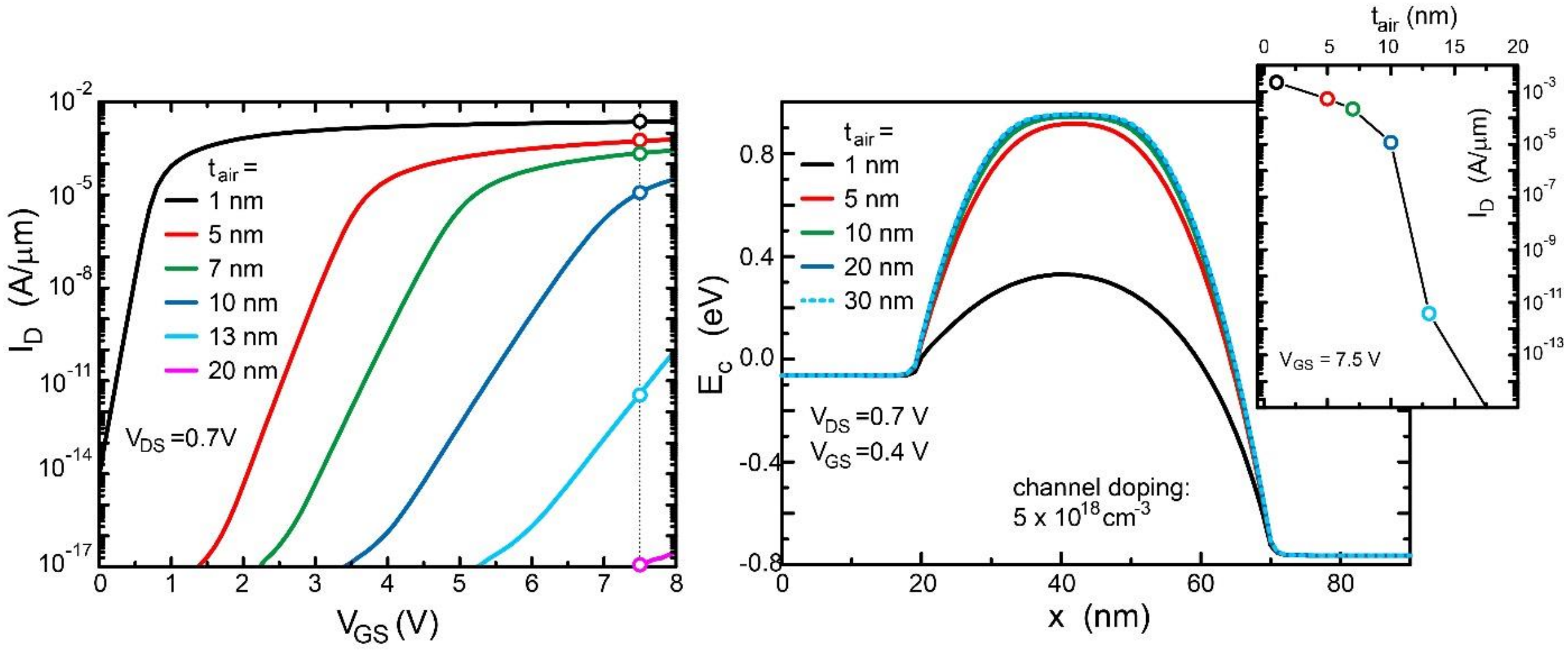

**Fig. 6.** Transfer characteristics (left) and conduction band profiles (right, at $V_{GS}$=0.4 V) of the model transistor, structure 3 with channel doping 5x$10^{18}$ cm$^{-3}$ (for a constant drain-source voltage of 0.7 V) for different air gap thicknesses $t_{air}$ in nm. The inset shows the drain current extracted at constant $V_{GS}$ = 7.5 V as a function of $t_{air}$.

Finally, Fig. 6 shows the transfer characteristics (left) and conduction band edge (right) for the MG MOSFET structure 3 with a rather high channel doping. In this case, short channel effects are strongly suppressed due to the doping as clearly observable in Fig. 6, right panel. However, for larger $t_{air}$, the doping (and thus the depletion capacitance) within the channel dominates, thereby degrading the inverse subthreshold slope for larger $t_{air}$ (see Fig. 6, left). Consequently, device operation as described for structure 1 and 2 does not work anymore. But a strong dependence of $I_D$ on $t_{air}$ can also be observed in structure 3, if the device is operated in its on-state, e.g. at $V_{GS}$=7.5 V as shown the inset of Fig. 6. The reason for this strong impact is the changing ratio between $C_{ox}$ ( $\propto 1/t_{air}$) and $C_{depl}$ as elaborated on below. Note, however, that in contrast to device structure 1 and 2, in structure 3 the highest drain current is obtained at smallest $t_{air}$, a behavior expected for a rather conventional device design as in Ref. [16].

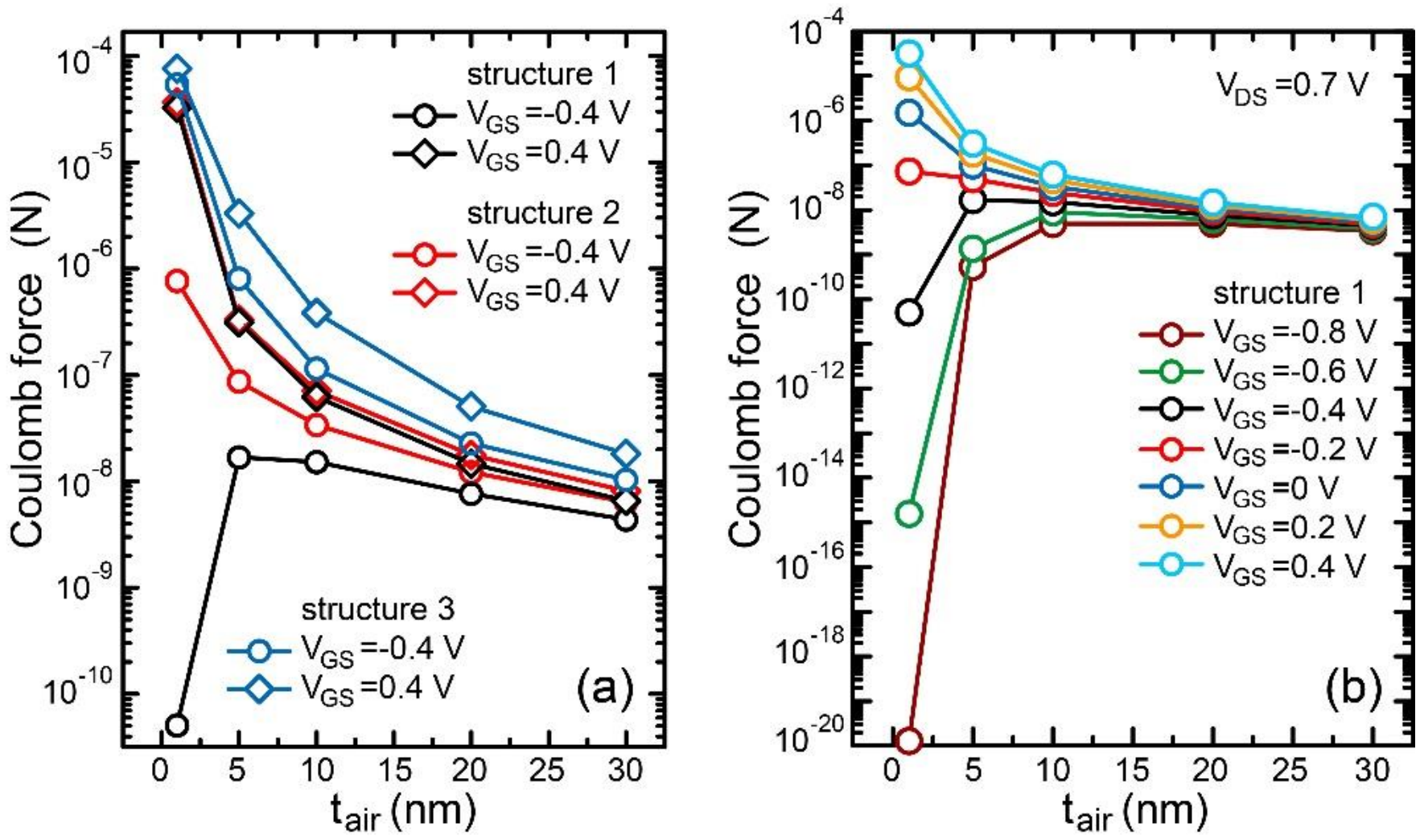

**Fig. 7.** Calculated Coulomb force vs air gap thickness for the model MG MOSFET. Left: Comparison of all three device structures for one positive/negative $V_{GS}$. Right: Coulomb force as a function of $t_{air}$ for structure 1 at different $V_{GS}$.

Next, in Fig. 7 the Coulomb forces for all three device structures calculated for different gate-source voltages are shown as a function of $t_{air}$, again for an applied drain-source voltage of 0.7 V. The left panel of Fig. 7 indicates that for structure 2 and 3 the Coulomb force increases continuously for decreasing $t_{air}$, regardless of the gate-source voltage. As shown in Eqn. (3), the Coulomb force in general is proportional to the square of the channel net charge and inversely proportional to the square of the distance between the net charge and the cantilever gate. Thus, for decreasing $t_{air}$ the denominator in (3) grows rapidly and this can only be compensated if at the same time the net charge decreases to a greater extent than the separation between net charge and gate decreases. If such compensation does not occur, and this is the case for structures 2 and 3, a pull-in of the cantilever will eventually occur. The situation is different for structure 1 as shown in the right panel of Fig. 7: Here, the Coulomb force increases to a lesser extent if the device is operated in its off-state, i.e. for $V_{GS} < -0.2$ V. Even more important, at larger negative $V_{GS}$, the Coulomb force drops rapidly for $t_{air} \leq 5$ nm, thereby avoiding a pull-in.

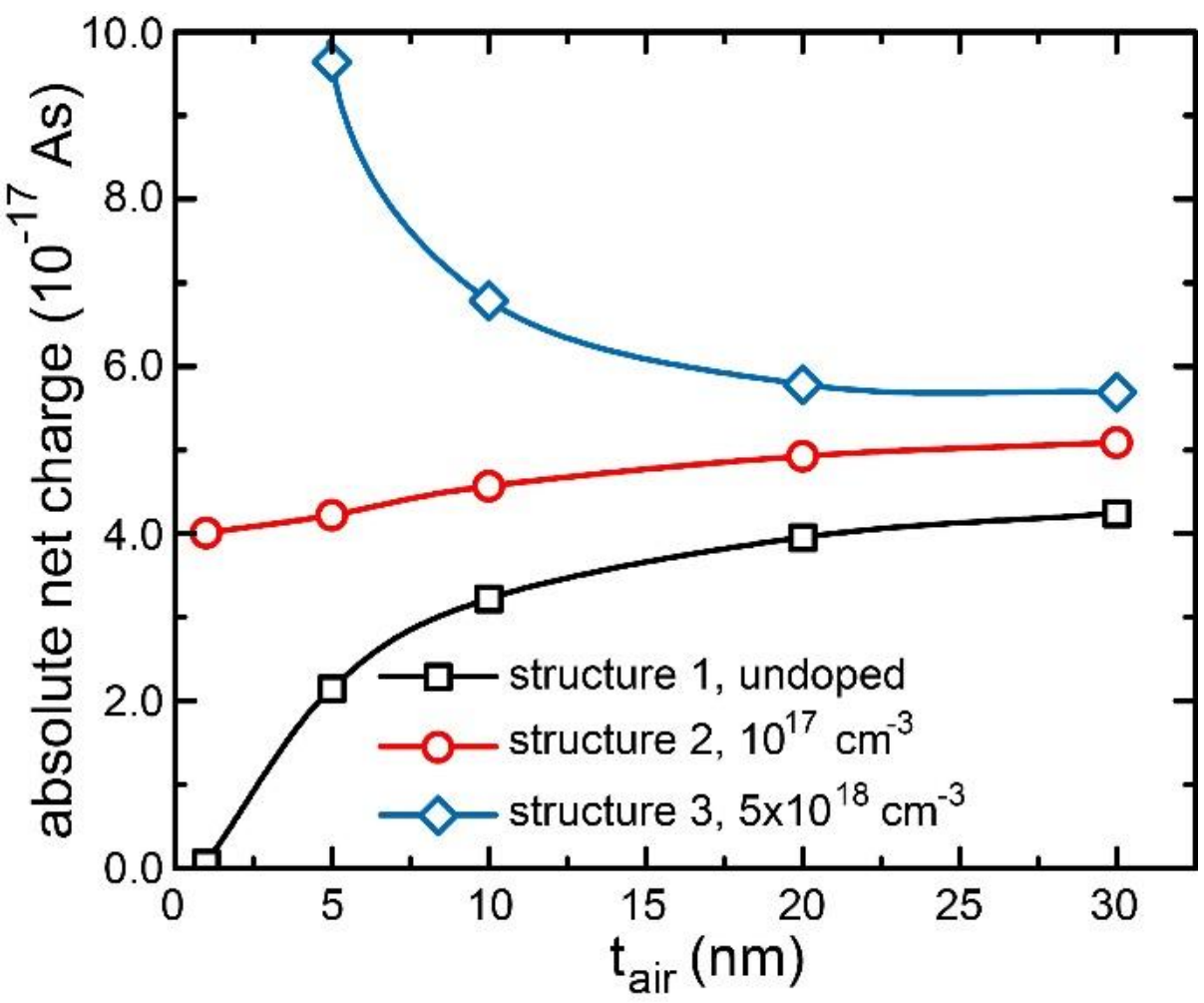

**Fig. 8.** Absolute value of the net charge versus air gap thickness of structures 1, 2 and 3 calculated for $V_{GS}$ = - 0.4 V and $V_{DS}$ = 0.7 V.

The reason for this behavior is revealed in Fig. 8 showing the absolute value of the net charge versus $t_{air}$ for all device structures for a gate-source voltage of -0.4 V. As can be seen, for structure 1 (black curve) the magnitude of the net charge $|Q_{net}|$ decreases continuously for decreasing $t_{air}$ due the reduction of SCE resulting in a removal of carriers (negative charge) from the channel. Below $t_{air}$ = 5 nm, this reduction of $|Q_{net}|$ overcompensates the effect of the decreasing $t_{air}$, thus leading to a diminishing Coulomb force. The magnitude of the net charge of structure 2 (red curve), on the other hand, reduces only moderately when $t_{air}$ is decreased, and this reduction is not sufficient to compensate the effect of the decreasing $t_{air}$ so that the Coulomb force increases continuously for decreasing air gap thickness. This effect is even more pronounced in the case of structure 3, where the Coulomb force strongly increases with decreasing $t_{air}$ (blue curve); the reason for the latter effect is the increasing ionization of acceptors (i.e. negative charge) in the channel when $t_{air}$ is reduced.

While our simulations only compute the static behavior of MG MOSFETs, the strong and non-monotonic dependence of the Coulomb force on $t_{air}$ suggests that it should be possible to excite the cantilever into self-sustained oscillations by merely applying appropriate static voltages. Note that in the present study, only the body doping was changed. However, there are many more degrees of freedom for tuning (see next section). For example, the cantilever gate dimensions $L_{cant}$ and $h_{cant}$ can be varied to tailor the mechanical cantilever properties, designs where the cantilever overlaps source and drain (i.e. larger than $L_G$) can be used to increase the level of Coulomb force to tune the excitation of cantilever oscillations, or the oxide thickness $t_{ox}$ and $L_G$ can be changed to adjust the short channel effects. In addition, different $V_{DS}$ allow adjusting the drain-induced barrier lowering and thus the on-state current obtained for large $t_{air}$. Moreover, the approach for calculating the cantilever deflection can be refined, e.g., by assuming not a constant Coulomb force but a $z$-dependent $F_C$ along the cantilever length.

### *C. Discussion of Simulation Results*

The results of the simulations discussed in the preceding section can be explained with a simple, so-called top-of-the-barrier model of MOSFET operation [24] which also allows setting up simple scaling rules for the MG MOSFET. Within the top-of-the-barrier model, the behavior of a MOSFET is reduced to the maximum potential barrier $\Phi_{max}$ (shown with the circles in Fig. 4 and 5, right panels) within the channel since this is responsible for carrier injection and hence the current through the device. The dependence of this maximum on the terminal voltages can be expressed through capacitive coupling of the source, gate and drain terminals to $\Phi_{max}$ according to [25]

$$\Phi_{max} = \frac{-e(V_{GS}-V_{BI})}{1+\varepsilon_{si}\frac{t_{air}\,t_{si}}{L_G^2}} + \frac{-eV_{DS}}{1+\frac{L_G^2}{\varepsilon_{si}t_{air}t_{si}}} + \frac{e^2 N_A t_{si}}{\varepsilon_0(\frac{1}{t_{air}}+\varepsilon_{si}\frac{t_{si}}{L_G^2})} \quad (4)$$

where $N_A$ is the ionized acceptor concentration and $V_{BI}$ is the built-in potential accounting for the doping level of the poly-Si gate electrode. Here, we neglected inversion charge in the channel and assume the depletion charge per area to be constant, i.e. $N_A t_{si}$, in order to simplify the analysis. With Eqn. (4), the different operating modes of the MG MOSFET and their dependence on $t_{air}$ and the channel doping can be explained. Furthermore, simple scaling rules for the design of MG MOSFETs can be extracted, too. In the case of zero to moderate doping (i.e. device structures 1 and 2), the third term in Eqn. (4) is zero or has a relatively small impact on $\Phi_{max}$ as observed when comparing the current voltage characteristics of device structure 1 and 2 (cf. Fig. 4 and 5, left panels). For device structure 1, Eqn. (4) yields a simple scaling rule by denoting $\varepsilon_{si}t_{air}t_{si}=\lambda^2$. In this case, the first term in Eqn. (4) reflects the loss of gate control due to SCE while the second term shows DIBL. The scaling factor $\lambda/L_G$ determines the amount of short channel effects and therefore the response of the MG MOSFET to a movement of the cantilever. In the present case, $\lambda/L_G$ varies between ~0.2 ($t_{air}$ = 1 nm) and 1.16 ($t_{air}$ = 30 nm). Comparing the change of $\Phi_{max}$ provided by Eqn. (4) with the simulated values extracted from Fig. 5 and 6 (right panels) one observes that Eqn. (4) overestimates the impact of SCE. The reason for this is that in the case of the simulation at $V_{GS}$ = -0.4 V, a significant inversion charge is present when $t_{air}$ = 30 nm that reduces the impact of SCE; extracting $\Phi_{max}$ at more negative $V_{GS}$ will lead to a better agreement. However, the qualitative behavior with an increase of $\Phi_{max}$ with reducing $t_{air}$ is well reproduced and shows that a MG MOSFET design with $\lambda/L_G$ ~1 when $t_{air} = t_{ox}$ yields a variation of the drain current over several orders of magnitude at constant $V_{GS}$ and $V_{DS}$.

In contrast to zero or moderate doping, for larger doping concentration the third term becomes important which leads to a reversed behavior: in this case, an increase of $t_{air}$ yields a stronger impact of the third term and hence an increase of $\Phi_{max}$ which leads to a reduction of the drain current. At the same time, the gate control over $\Phi_{max}$ is strongly reduced (see Fig. 6, right panel) and as a result, a device with substantial channel doping (i.e. structure 3) needs to be operated in the on-state as has been shown in the preceding section (cf. Fig. 6). While the on-state operation of structure 3 also enables a very large current change with moving cantilever (see inset Fig. 6) it inevitably leads to a pull-in due to an increasing Coulomb force (cf. Fig. 7 (a) and Fig. 8). In general, the presence of dopants in the channel leads to an increasing Coulomb force and thus to a pull-in when $t_{air}$ becomes small even in the case of moderate doping. Therefore, undoped channels are best suited for our MG MOSFETs.

Summarizing the discussion, the major difference between device structures1, 2 and structure 3 is that in the latter case, the drain current *increases* with decreasing $t_{air}$ since the device is operated in its on-state; note that the required, high gate voltage $V_{GS}$=7.5 V is due to the large threshold voltage shift. When $t_{air}$ is decreased the gate impact increases yielding a larger current increase but also a large increase of the channel charge resulting in a pull in (see Fig. 5). In contrast, the device structures 1 and 2 are operated in the off-state and hence, the drain current *decreases* with decreasing $t_{air}$. In the case of structure 1 the reduction of SCE removes the charge in the channel and hence avoids pull-in. At the same time, for large $t_{air}$, the current becomes rather large due to drain-induced barrier lowering. These are the two main features of our cantilever design that enable their use in the sensor system discussed in section 1 (see Fig. 1). It has already been mentioned above that the strong change of current as a function of cantilever position can be exploited to connect the cantilever sensors directly to a logic gate. To this end, the cantilever has to be driven into self-oscillation by a suitable actuation [26, 27]. Our theoretical analysis above suggests that this could be accomplished by simply applying appropriate voltages at gate and drain and exploiting the strong non-monotonic dependence of the Coulomb force on the cantilever position (see Fig. 7). In this case, the channel acts as actuator electrode with variable Coulomb force due to the modulate SCE.

## 3. Experimental Demonstration of MG MOSFETs

### *A. Cantilever Fabrication*

In order to demonstrate the functionality of our MG MOSFET concept to enable a highly sensitive read-out mechanism, we fabricated such a device. To facilitate observing and measuring the distance between the cantilever gate and the channel of the MOSFET, the device was rotated by 90° and a lateral architecture as illustrated in Fig. 9 (a) was implemented. Apart from the easy observability of a lateral cantilever movement, an additional benefit of this layout is that the devices are based on SOI material where all relevant device parameters can be adjusted with lithography and silicon etching. For instance, the stiffness of the cantilever can be adjusted by choosing different thicknesses $h_{cant}$ and/or cantilever lengths $L_{cant}$; the parameters of the integrated MOSFET can be modified by adjusting the air gap thickness $t_{air}$, the channel layer thickness $t_{si}$ (equivalent to the SOI thickness in Fig. 2) and the channel length $L_G$ (cf. Fig. 9 (b)). Moreover, additional electrostatic actuator electrodes can be realized since all parts of the device reside on BOX and are thus insulated from each other.

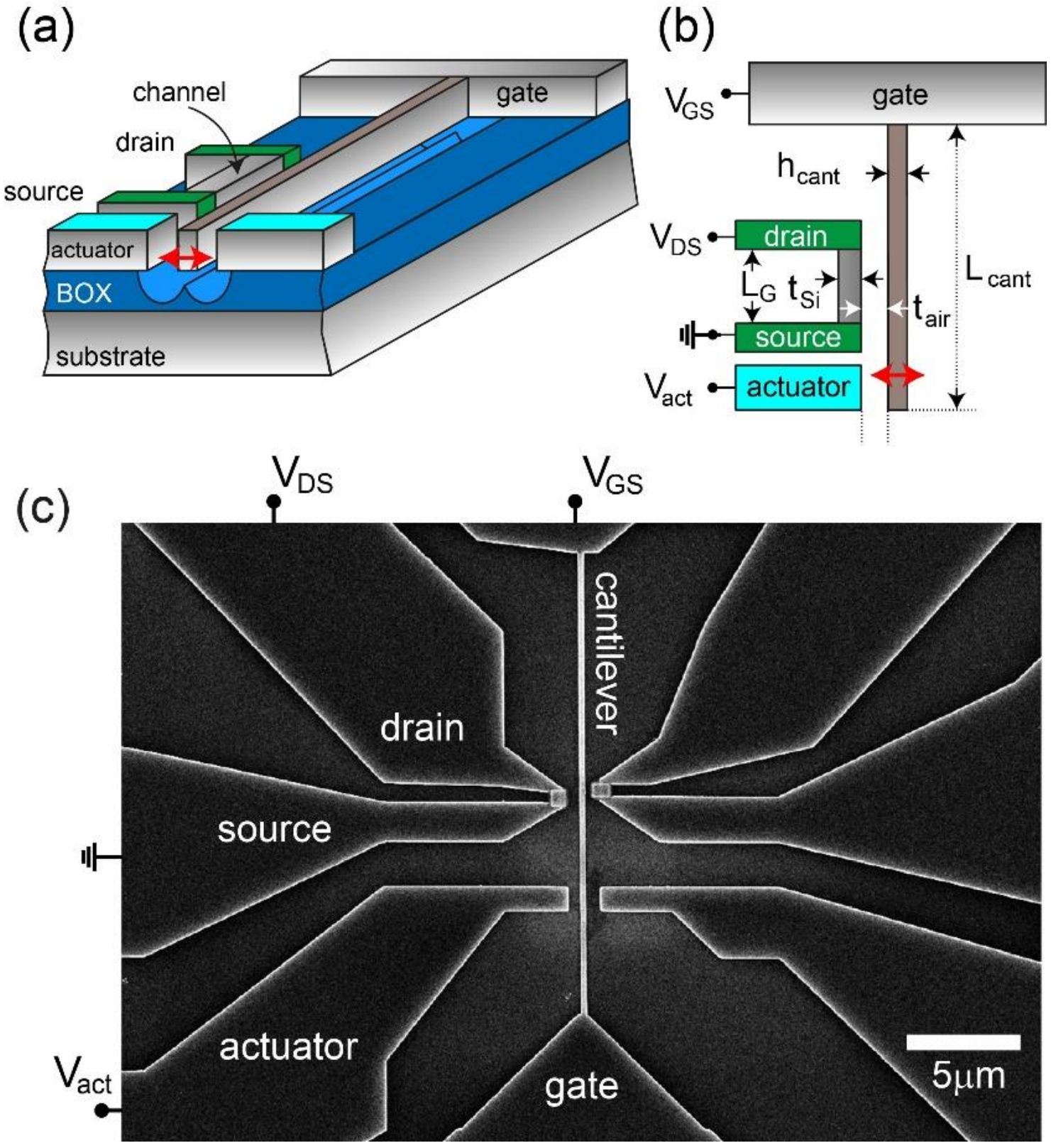

**Fig. 9.** Illustration of a fabricated MG MOSFET. (a) and (b): Laterally actuated cantilever MG MOSFET design. (c) Electron micrograph of a fabricated lateral cantilever MG MOSFET with additional electrostatic actuators.

For the device fabrication, SOI with a top silicon layer of 205 nm (p-doped, resistivity 13.5 - 22.5 Ω cm) and a BOX thickness of 200 nm was used. Electron-beam lithography (EBL) was applied together with PMMA (950k, 3% solid compound, 160 nm) to generate a mask pattern for a subsequent etching in a deep silicon etcher. A cryo etching process at 143 K with a gas mixture consisting of 37.5 sccm $SF_6$ and 14 sccm $O_2$ was used to etch through the top silicon layer, stopping on the buried oxide. After the removal of the PMMA mask in an $O_2$ plasma for 60 mins, a second EBL step is used where the channel area of the FET (cf. Fig. 9 (c)) is covered. In order to reduce exposure time, we employed a negative tone EBL resist for covering the areas not to be implanted, consisting of diluted AZ nLOF2070. The final resist thickness of 300 nm was sufficient to prevent doping the channel during ion implantation, which could be verified using sheet resistance measurements. Arsenic was implanted at an energy of 40 kV and an area dose of $2 \times 10^{15}$ $cm^{-2}$, using a 7° sample tilt w.r.t. the direction of incident ions. The implantation was followed by an activation anneal at 800 K for 60 s in argon atmosphere in a rapid thermal annealing tool. Subsequently, the BOX was removed to detach the patterned structures enabling their free movement, using hydrofluoric acid (HF, 50%) for 25 s. Finally, a rinse in DI water followed by a drying step in boiling isopropanol was employed to avoid any residually trapped liquid to cause mutual sticking of the detached silicon structures.

Figure 9 (c) shows an electron micrograph of a cantilever alongside two MOSFET devices and two actuator electrodes. Within the channel area of the two MOSFETs a resist mask patterned with EBL is visible that is used to protect the channel during the As implantation of the source, drain gate and actuator electrodes.

### *B. Measurement Results and Discussion*

Figure 10 (a) shows the transfer characteristics of the fabricated cantilever MOSFETs for four different $V_{DS}$. A voltage of $V_{act} = -80$ V was applied at the actuator gate. For gate voltages less than approximately 40 V a rather constant and $V_{GS}$-independent leakage current is obtained whose magnitude depends on $V_{DS}$. We attribute this large leakage to strong short-channel effects. Together with a rather large air gap thickness of ~ 600 nm there is hardly any gate impact and the current solely depends on $V_{DS}$. However, with increasing gate voltage the voltage between actuator and gate also increases and the cantilever is moved towards the channel of the MOSFET, effectively reducing $t_{air}$. As a result, short-channel effects are reduced and the gate impact increases, eventually switching the transistor on, yielding a ratio between on- and off-current of more than seven orders of magnitude (for $V_{DS} = 0.5$ V). For the operation of the cantilever MG MOSFET we chose a gate voltage of $V_{GS} = 50$ V and measured the drain current as a function of the actuator voltage $V_{act}$. Figure 10 (b) displays the results showing that $I_D$ can be changed by approximately five orders of magnitude in the case of the smallest bias simply by changing the actuator voltage between 0 V and -80 V, i.e., effectively between -50 V and -130 V due to the static cantilever potential (= gate potential) of 50 V. Since a non-zero $V_{act}$ leads to an electrostatic attraction of the cantilever, i.e., a reduced gate dielectric thickness $t_{air}$, and since the current increases with larger $V_{act}$, the device operates in the on-state.

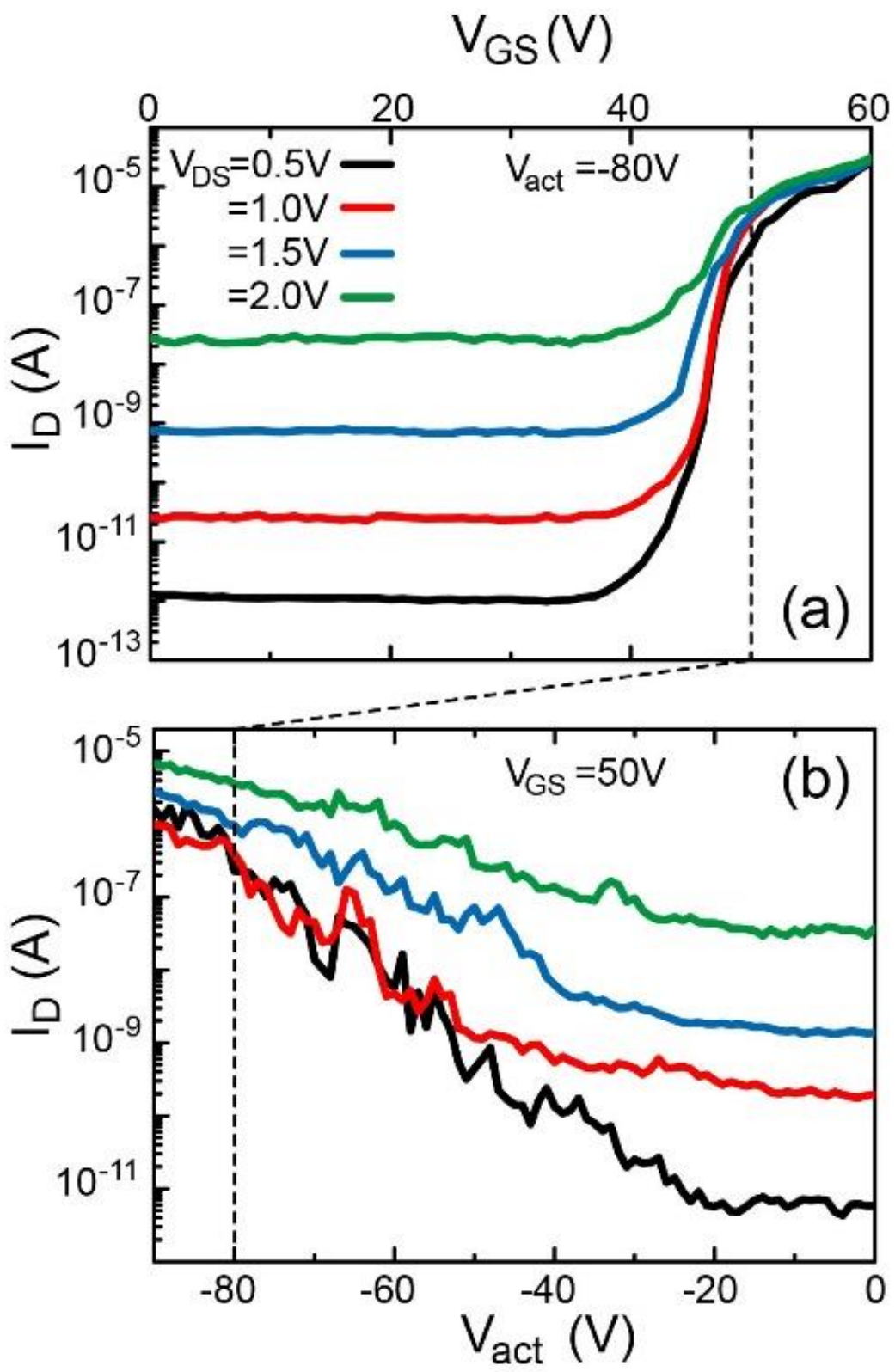

**Fig. 10**. Measured current-voltage characteristics of the fabricated cantilever MG MOSFET. (a) Transfer characteristics of the cantilever FET for different $V_{DS}$ at an actuator voltage of $V_{act} = -80$ V. At sufficiently large effective actuator voltage $V_{act}$ - $V_{GS}$ the cantilever is moved towards the transistor resulting in a stronger gate impact with less short-channel effects. (b) Drain current as a function of $V_{act}$ at a constant $V_{GS} = 50$ V.

In order to transfer Fig. 10 (b) into a dependence of $I_D$ on $t_{air}$, we measured the distance between cantilever and MOSFET channel within a scanning electron microscope while applying the appropriate $V_{GS}$ and $V_{act}$, respectively; the resulting dependence is displayed in the inset of Fig. 11. Despite a measurable contribution of the negative charges imparted by the SEM during the measurement, the magnitude of the effect was negligible and resulted in an excess current flow of no more than $10^{-10}$ A. Since all areas that had to be scanned for the measurement were equally affected by the electron beam, the only voltage shift to be observed was a constant offset voltage against ground, leaving all relative potentials determining the device functionality unaffected.

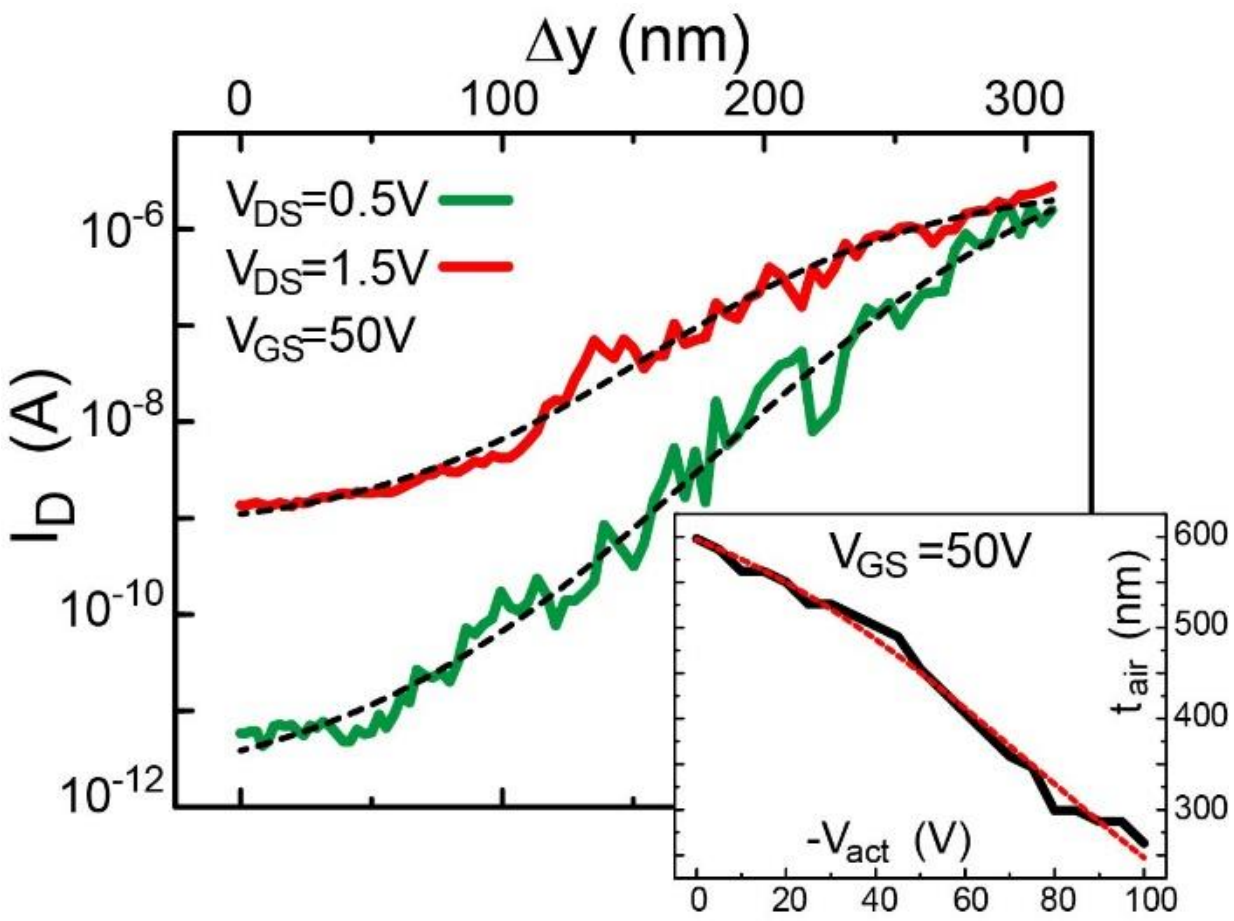

**Fig. 11.** Drain current versus change of the position of the cantilever. Moving the cantilever 300 nm closer to the channel of the MOSFET results in six orders of magnitude higher current flow. The inset shows the distance between cantilever and channel as a function of $V_{act}$ measured with scanning electron microscopy. The dashed lines are fits to the experimental data.

The main panel of Fig. 11 displays the current $I_D$ at a constant $V_{GS}$ as a function of the change of the cantilever position $\Delta y$. It shows that a modulation of the drain-current over several orders of magnitude is indeed possible when moving the cantilever, i.e., the gate electrode. The results of our experimental work displayed in Fig. 11 qualitatively agree with our simulation work (see section 2) and clearly prove, that our concept of exploiting short-channel effects as a read-out mechanism provides several orders of magnitude current difference depending on the cantilever position.

To facilitate a direct integration of the cantilevers into a sensor as depicted in Fig. 1, the cantilever has to be stimulated to a self-sustained oscillation by a suitable actuation [26, 27]. Our theoretical analysis above suggests that this could be accomplished by simply applying appropriate voltages at gate and drain and exploiting the strong non-monotonic dependence of the Coulomb force on the cantilever position (see Fig. 7 (b)). In this case, the channel acts as actuator electrode with variable Coulomb force due to the modulation of SCE: whenever $t_{air}$ becomes less than ~5 nm, charge is removed from the channel and the Coulomb force breaks down resulting in an oscillation. Such an oscillation, induced by merely applying constant voltages, has already been shown in Fig. 1 (c) (note that the cantilever was fabricated in the same way as described in the preceding section). In the case of this cantilever, its resistance (and that of the actuator electrode) can be adjusted such that when the cantilever is close enough to the actuator electrode, charge is transferred from the cantilever to the actuator electrode. With appropriately high resistances, this loss of charge cannot be compensated quickly enough, leading to a breakdown of the Coulomb force causing the cantilever to swing back. While the exact mechanism of the Coulomb force breakdown is different in the cantilever shown in Fig. 1 (c) and the MG MOSFET, in both cases the Coulomb force breaks down because the charge responsible for it is removed when the cantilever approaches the counter electrode (actuator or FET, respectively). No pull-in occurs although a constant voltage has been applied. Therefore, we expect a selfsustained oscillation to also occur in properly designed MG MOSFETs.

## 4. Conclusion

We simulated and fabricated functional devices proving the feasibility of utilizing a movable gate MOSFET with a direct electrical readout mechanism based on short-channel effects for sensor applications. This MG MOSFETs provides several orders of magnitude transduction of mechanical movement into an electrical signal. When operated in the off-state, pull in is avoided due to a strong, non-monotonic dependence of the Coulomb force on the cantilever movement. Finally, if such cantilevers are excited into a self-sustained oscillation with merely applying appropriate DC voltages, a monolithically integrable and up-scalable nano-electro-mechanical sensor can be realized that allows direct connection to the inputs of a logic AND gate avoiding the need for analog to digital signal conversion.

## Acknowledgments

Financial support by the Bosch Forschungsstiftung and the Federal Ministry of Education and Research of Germany (BMBF) under grant no. 16ES0060K is gratefully acknowledged.